\documentclass[reprint, amsmath, amssymb, aps]{revtex4-1}%
\pdfoutput=1
\usepackage{hyperref}
\usepackage{graphicx}
\usepackage{dcolumn}
\usepackage{amsmath}
\usepackage{amsfonts}
\usepackage{amssymb}
\usepackage{xcolor}
\usepackage{ascmac}
\usepackage{listings}
\begin{document}

\title{SWRC Fit and unsatfit for parameter determination of unsaturated soil properties}
\thanks{Published at \textit{Journal of Toyo University Natural Science}, https://doi.org/10.34428/0002000817}

\author{Katsutoshi Seki}
\affiliation{Natural Science Laboratory, Toyo University\\
5-28-20 Hakusan, Bunkyo-ku, Tokyo 112-8606, Japan\\
\href{https://orcid.org/0000-0002-5707-0283}{orcid.org/0000-0002-5707-0283}}
\begin{abstract}

SWRC Fit and unsatfit are programs developed for determining parameters of the water retention and the unsaturated hydraulic conductivity functions of soils for analyzing water movement in the unsaturated soil. The SWRC Fit is a web application for drawing SWRC (soil water retention curves) of various soil hydraulic models from the measured data with just one click, available at https://purl.org/net/swrc/. The SWRC Fit depends on the Python library unsatfit that implements various types of soil hydraulic models to determine water retention and hydraulic conductivity parameters by nonlinear least-square optimization. The implemented models are classified as DF3 (Brooks and Corey, van Genuchten, Kosugi) models, DF4 (Fredlund, Fayer, Peters, bimodal-CH) models, and DF5 (bimodal) models by the degree of freedom of the water retention functions. Most of the hydraulic conductivity functions are derived from the water retention function and the general hydraulic conductivity function. A modified model to suppress extreme changes in hydraulic conductivity in the near-saturated range can be applied to any hydraulic model. The algorithm that was proved to work with various types of soils in Seki et al. (2023) is provided as the sample Python codes. This paper describes how to use SWRC Fit and unsatfit.

\end{abstract}%

\maketitle
\section{\label{sec:intro}Introduction}

For analyzing soil water movement in the vadose zone, the Richards equation (Ref. \cite{richards1931}) is solved numerically (Ref. \cite{farthing2017}). One-dimensional vertical water flow without sink term is expressed with Richards equation as

\begin{equation}
\frac{\partial \theta}{\partial t}= \frac{\partial}{\partial z}\biggl[K\left(\frac{\partial h}{\partial z} + 1 \right)\biggr]
\label{eq:richards}
\end{equation}

\noindent where $h$ is the pressure head [L], $\theta$ is the volumetric water content [$\mathrm{L}^3\mathrm{L}^{-3}$], $t$ is time [T], $z$ is the coordinate positive upward [L], and $K$ is the unsaturated hydraulic conductivity [LT$^{-1}$]. The equation can be solved when the unsaturated soil properties in the flow domain are defined by water retention function (WRF) $\theta(h)$ and hydraulic conductivity function (HCF) $K(h)$, and initial and boundary conditions are given. Note that in equation \ref{eq:richards} $h$ is defined as negative for unsaturated conditions, but after this paragraph, $h$ is defined as positive for the unsaturated condition for notational convenience.

\begin{figure}
	\includegraphics[scale=0.8]{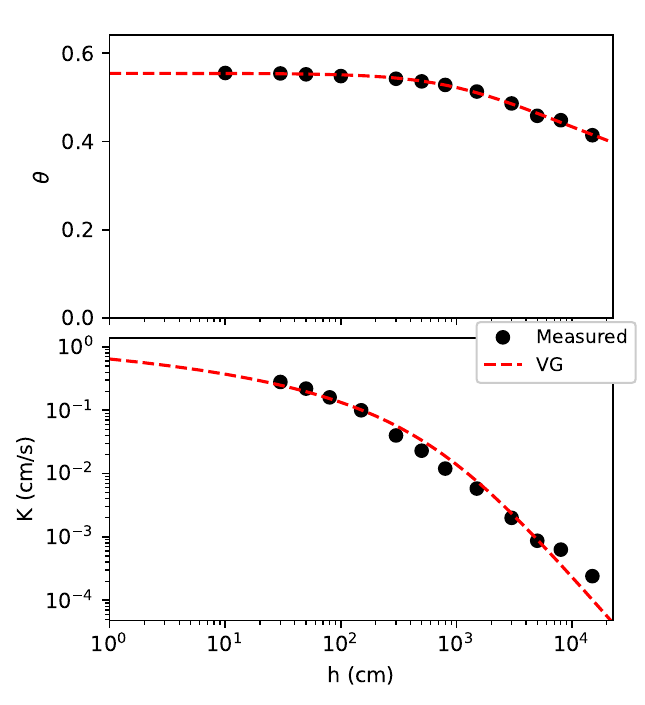}
	\caption{\label{fig:unsoda2363}Water retention and hydraulic conductivity curves of clay soil of UNSODA 2363.
	Measured data and fitted curves with the van Genuchten - Mualem model are shown.}
\end{figure}

Fig. \ref{fig:unsoda2363} shows water retention and hydraulic conductivity data of clay soil from the UNSODA database (Ref. \cite{nemes2001}). As shown in this figure, when we have a measured dataset of ($h$, $\theta$), we can get a fitted water retention curve to the measured data with an appropriate WRF; in this case, van Genuchten equation (Ref. \cite{vg1980}). The WRF has several parameters, (WRF parameters) and they are optimized with the curve-fitting. In the same way, when we have a measured dataset of ($h$, $K$), we can fit the hydraulic conductivity curve to the measured data with an appropriate HCF; Mualem equation (Refs. \cite{mualem1976, vg1980}) is used in Fig. \ref{fig:unsoda2363}. As we already got the WRF parameters from the water retention data, we determine the HCF parameters from ($h$, $K$) data by a nonlinear optimization method; here we define the HCF parameters as the parameters in HCF which are not in the WRF parameters.

In reality, the WRF and the HCF parameters can be determined by a flow simulation with the Richards equation to minimize the error of the simulated and measured $h$, $\theta$, and/or water flux in a controlled laboratory experiment, for example by a multistep outflow method (Ref. \cite{dam1994}). In this case, it is still preferable to have the WRF parameters determined from the measured ($h$, $\theta$) (Ref. \cite{rudiyanto2013}) because there are so many parameters and the optimization is easier with smaller numbers of unknown parameters.

The SWRC Fit is a program for obtaining WRF parameters of various hydraulic models from a given set of ($h$, $\theta$) data with a simple web interface. A Python library unsatfit is for fitting the WRF and the HCF parameters from a given set of ($h$, $\theta$) and ($h$, $K$) data. This paper describes how to use SWRC Fit and unsatfit.
\section{\label{sec:model}Models}

\subsection{Water retention functions}

\subsubsection{Basic models}

The effective saturation $S$ as a function of the pressure head $h$ is defined as
\begin{equation}
	S(h) = \frac{\theta(h)-\theta_r}{\theta_s - \theta_r}
\end{equation}
where $\theta_s$ and $\theta_r$ are the saturated and residual volumetric water content [$\mathrm{L}^3\mathrm{L}^{-3}$], respectively.
Therefore WRF $\theta(h)$ can be expressed in terms of $S(h)$, and WRF parameters are $\theta_s$, $\theta_r$, and parameters in $S(h)$ function. Note that the pressure head $h$ can also be expressed in terms of pressure [M L$^{-1}$ T$^{-2}$], but in this paper, we use [L] for the pressure head.

The Basic models (BC, VG, KO, and FX models) are as follows.
The Brooks and Corey (BC) model is defined as (Ref. \cite{brooks1964})
\begin{equation}
	S(h) = \begin{cases}\left(h / h_b\right)^{-\lambda} & (h>h_b) \\ 1 & (h \le h_b)\end{cases}
	\label{eq:bc}
\end{equation}
where $h_b$ [L] and $\lambda$ are parameters.

The van Genuchten (VG) model is defined as (Ref. \cite{vg1980})
\begin{subequations}
	\label{eq:vg}
	\begin{equation}
		S(h) = \biggl[\dfrac{1}{1+(\alpha h)^n}\biggr]^m
	\end{equation}
	\begin{equation}
		m=1-q/n
	\end{equation}
\end{subequations}

\noindent where $\alpha$ [L$^{-1}$], $m$, $n$ are parameters. $m$ and $n$ are bounded by $0<m<1$ and $n>q$. Usually $q=1$ is assumed and therefore $n>1$, and the free variables are only $\alpha$ and $n$ (or $\alpha$ and $m$).

The Kosugi (KO) model is defined as (Ref. \cite{kosugi1996})
\begin{equation}
	S(h) = Q \biggl[\dfrac{\ln(h/h_m)}{\sigma}\biggr]
	\label{eq:ko}
\end{equation}
where $h_m$ [L] and $\sigma$ are parameters and $Q(x)$ is a complementary cumulative normal distribution function defined by
\begin{eqnarray}
	Q(x) &=& \int_{x}^{\infty}\frac{\exp(-x^2/2)}{\sqrt{2\pi}}dx\nonumber\\\
	&=& \mathrm{erfc}(x/\sqrt{2})/2
\end{eqnarray}

The Fredlund und Xing (FX) model is defined as (Ref. \cite{fredlund1994})

\begin{eqnarray}
	S(h) = \biggl[ \dfrac{1}{\ln \left[e+(h / a)^n \right]} \biggr]^m
	\label{eq:fx}
\end{eqnarray}
where $e$ is Napier's constant and $a$, $m$, $n$ are parameters.

\subsubsection{Bimodal and bimodal-CH models}

For adding flexibility to the water retention curve, especially for fitting a wide range of pressure heads, the multimodal WRF is defined as (Refs. \cite{durner1994, seki2022})
\begin{equation}
	S(h) = \sum_{i=1}^{k}w_iS_i(h)
\end{equation}
where $k$ is the number of subsystems, and $w_i$ are weighting factors with $0 < w_i < 1$ and $\sum w_i = 1$, and $S_i(h)$ is a subfunction that is equivalent to one of the BC (equation \ref{eq:bc}), VG (equation \ref{eq:vg}) and KO (equation \ref{eq:ko}) models. The combined model is denoted by a subscript of the number of the subfunction. For example, the KO$_1$BC$_2$ model denotes KO-type for $S_1(h)$ and BC-type for $S_2(h)$. Combinations of the same subfunctions (e.g., BC$_1$BC$_2$BC$_3$) are referred to as multimodels (e.g., multi-BC). Multimodels consisting of only two similar subfunctions are referred to as dual-models, such as the dual-VG for VG$_1$VG$_2$. As the multimodels have parameters of the same name for the different subsystems, parameters are distinguished by the subscript of the number of the subfunctions. For example, dual-VG model has $\alpha_1$, $n_1$ for $S_1(h)$ and $\alpha_2$, $n_2$ for $S_2(h)$. In multimodels, usually, $\theta_r = 0$ is assumed.

The CH (common head) model (Ref. \cite{seki2022}) assumes that ${h_b}_i$ for the BC model, $\alpha_i^{-1}$ for the VG model, and ${h_m}_i$ for the KO model has the same value $H$, i.e., $H = {h_b}_i = \alpha_i^{-1} = {h_m}_i$. We have, for example, the KO$_1$BC$_2$-CH model, dual-VG-CH model, etc. While the dual-VG model has 6 parameters ($\theta_s$, $w_1$, $\alpha_1$, $n_1$, $\alpha_2$, $n_2$), the dual-VG-CH model has 5 parameters ($\theta_s$, $w_1$, $H$, $n_1$, $n_2$). The bimodal model ($k=2$) with the CH assumption is denoted as the bimodal-CH model.

Several bimodal models are implemented in unsatfit as listed on the "Models" page on the website of the unsatfit (Ref. \cite{unsatfit}). The models implemented in the SWRC Fit are listed on the website of the SWRC Fit (Ref. \cite{swrcfit}, follow the link of "several soil hydraulic models").

\subsubsection{Models of fixed oven dry pressure}

The unsatfit implements some more models, such as Fayer and Simmons (FS) model \cite{fayer1995} (VG-type equation) and Peters (PE) model \cite{peters2013} (KO-type equation). Both models have 2 more variables in addition to the basic (VG, KO) models. The first parameter is comparable to $w_1$ in the bimodal model. The second parameter $h_e$ is a pressure head at the oven-dry condition, where $\theta$ first becomes zero. Therefore they are similar to bimodal models, where the second subfunction has a fixed pressure head of the $\theta=0$ point condition.

\subsection{Hydraulic conductivity functions}

\subsubsection{\label{sec:hcmodel}General HCF}

The general HCF is defined as (Ref. \cite{hoffmann1999})
\begin{equation}
	{K_{\textrm{r}}}{\textrm{\ }}\left( h \right) = \frac{{K\left( h \right)}}{{{K_{\textrm{s}}}}}\ = S{\left( h \right)^p}\ {\left[ {\frac{{\mathop \int \nolimits_0^{S\left( h \right)} h{{\left( S \right)}^{- q}}{\textrm{d}}S\ }}{{\mathop \int \nolimits_0^1 h{{\left( S \right)}^{ - q}}{\textrm{d}}S\ }}\ } \right]^r}
	\label{eq:ghc}
\end{equation}
where $K_s$ is the saturated hydraulic conductivity, $K_r$ is the relative hydraulic conductivity, and where $p$, $q$, and $r$ are HCF parameters. Note that $q$ is common with equation \ref{eq:vg} of the VG model.

As the HCF includes the integration of the function $h(S)$, it is convenient when a closed-form expression of the integrated function is obtained for a specified WRF; otherwise, numerical integration or approximation is required. Most of the HCF in the unsatfit are closed-form expressions of the general HCF derived from the respective WRF, as derived in Refs. \cite{priesack2006, seki2022}. The PE model is an exception.

The general HCF expresses different types of models with the HCF parameters, $p$, $q$, $r$. The Burdine model is for $p=2$, $q=2$, $r=1$ (Ref. \cite{burdine1953}), and the Mualem model is for $p=0.5$, $q=1$, $r=2$ (Ref. \cite{mualem1976}). When $p$ is used as a variable and changed from the original value in those models, $p$ is called a tortuosity factor.

Although normally ($q$, $r$) is fixed and $p$ optimized, it was shown that two parameters ($p$, $q$) or ($p$, $r$) should be optimized to represent the hydraulic conductivity curves over a wide range of pressure heads for the multimodal models (Ref. \cite{seki2023}). Specifically, the dual-BC-CH model with $r=1$ and optimization of ($p$, $q$), the dual-VG-CH model with $q=1$ and optimization of ($p$, $r$), and the KO$_1$BC$_2$-CH model with $r=1$ and optimization of ($p$, $q$) was evaluated with a broad range of soil types with the UNSODA database and it was shown that the models are useful for practical applications while mathematically being simple and consistent.

\subsubsection{\label{sec:modified}Modified model}

Since the VG model shows unrealistically large reductions in the hydraulic conductivity near saturation when $n$ approaches its lower limit of 1, a modified model which introduces a hypothetical air-entry head $h_b$ near saturation (where $S(h_b) \approx 1$) was developed (Ref. \cite{vogel2000}). The modified model was extended to the multimodal models (Ref. \cite{seki2022}). It can be generalized to any hydraulic model as follows.

By denoting the original WRF and HCF as $S(h)$ and $K_r(h)$, and the modified WRF and HCF as $S'(h)$ and $K_r'(h)$, the modified model can be described as
\begin{subequations}
\label{eq:modified}
\begin{equation}
S'(h) = \begin{cases}\dfrac{S(h)}{S(h_b)} & (h > h_b)\\
1 & (h \le h_b)\end{cases}
\end{equation}
\begin{equation}
K_r'(h) = \begin{cases}\dfrac{K_r(h)}{K_r(h_b)} & (h > h_b)\\
1 & (h \le h_b)\end{cases}
\end{equation}
\end{subequations}

This definition matches those of Refs. \cite{vogel2000, seki2022}. As $S(h_b) \approx 1$ is assumed, $S'(h)$ is almost equal to $S(h)$. With this definition, we can make modified models for any kind of WRF and HCF.

\subsection{\label{selectmodel}Selection of a model}

The selection of the model depends on the purpose. If the purpose is only to describe the water retention curve, we can just perform the fitting of multiple WRFs simultaneously with the SWRC Fit as described later and decide the model by looking at the fitting performance such as $R^2$, AIC, and fitted curves.

The degree of freedom of the WRF function is classified in Table \ref{table:df}. As $\theta_s$ can be measured and fixed, $\theta_s$ is not counted. $\theta_r$ is counted for the basic models, but not counted for other models where usually $\theta_r = 0$ is assumed. For the FS and the PE models, $h_e$ could be fixed, but it was counted as it can be a fitting parameter with some uncertainty (Ref. \cite{fayer1995}).

\begin{figure}
	\includegraphics[scale=0.8]{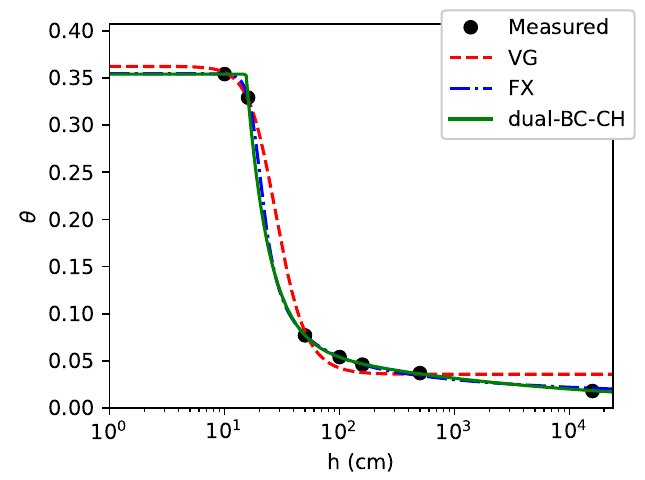}
	\caption{\label{fig:unsoda3332}Water retention curve of sand of UNSODA 3332 fitted by the VG, FX, dual-BC-CH models, where $\theta_r=0$ for dual-BC-CH only.}
\end{figure}

In most cases, the DF3 model is sufficient for WRF fitting (Fig. \ref{fig:unsoda2363}), while the DF4 model may be required for HCF fitting as will be discussed later. The FX model is better than the VG model for UNSODA 3332 (Ref. \cite{seki2017}). This is a typical soil requiring the DF4 model for WRF fitting, where the DF4 models (FX, dual-BC-CH) are much better than the DF3 (VG) models as shown in Fig. \ref{fig:unsoda3332}.

\begin{figure}
	\includegraphics[scale=0.8]{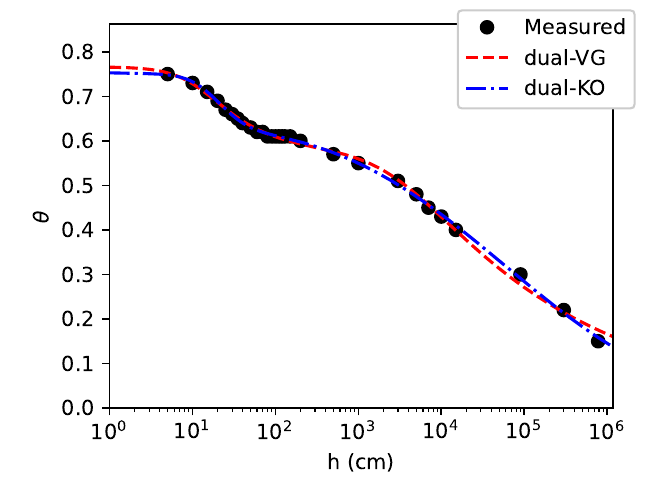}
	\caption{\label{fig:tachikawa}Water retention curve of Tachikawa Andisols (Ref. \cite{hamamoto2009}) fitted with the dual-VG and the dual-KO models where $\theta_r=0$.}
\end{figure}

For soils that have bimodal pore structures, the DF5 model is required. An example is Tachikawa Andisols with highly aggregated pore structure (Ref. \cite{hamamoto2009}) (Fig. \ref{fig:tachikawa}). As discussed in Ref. \cite{hamamoto2009}, the dual-KO model directly shows 2 peaks in the pore-size density function, as indicated by the parameters $h_{m1}$ and $h_{m2}$ corresponding to the inflection points in the water retention curve. Such bimodality cannot be expressed with the DF4 models, despite the name "bimodal-CH".

\begin{table}[h]
	\caption{Water retention functions classified by the degree of freedom (DF). $\theta_s$ is not counted. $\theta_r$ is counted only for the BC, VG, KO, and FX models. See section \ref{sec:model} for the model description. Note that $q=1$ is assumed for the VG model.}
	\label{table:df}
	\centering
	 \begin{tabular}{clll}
	  \hline
	  DF & Model \\
	  \hline 
	  DF3 & BC, VG, KO models\\
	  DF4 & FX, FS, PE, Bimodal-CH models\\
	  DF5 & Bimodal models \\
	  \hline
	 \end{tabular}
   \end{table}

\begin{figure}
	\includegraphics[scale=0.65]{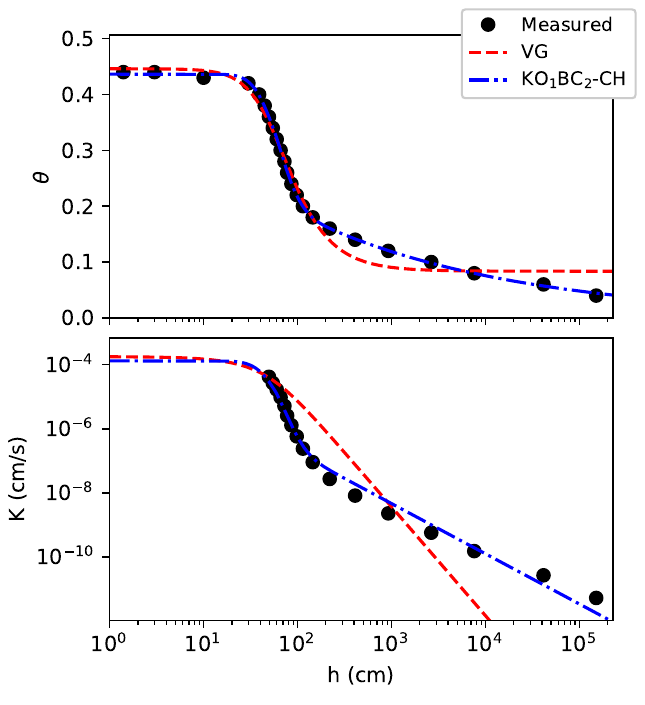}
	\caption{\label{fig:gilat}Water retention and hydraulic conductivity curves of Gilat loam fitted with the van-Genuchten Mualem model and the KO$_1$BC$_2$-CH model ($\theta_r=0$) with general HCF ($r=1$).}
\end{figure}

\begin{figure}
	\includegraphics[scale=0.65]{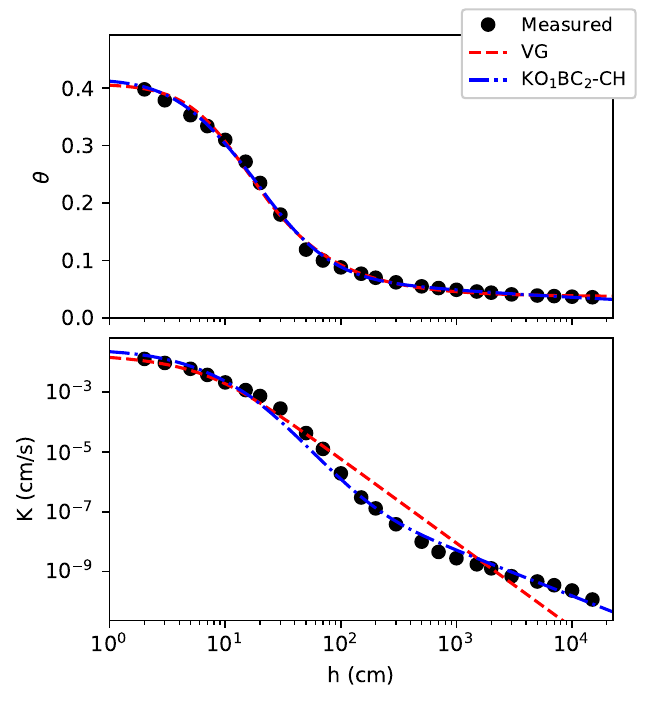}
	\caption{\label{fig:unsoda4661}Water retention and hydraulic conductivity curves of sand of UNSODA 4661 fitted with the van-Genuchten Mualem model and the KO$_1$BC$_2$-CH model ($\theta_r=0$) with general HCF ($r=1$).}
\end{figure}

If the purpose of selecting the model is to conduct the water flow simulation with Richards equation (equation \ref{eq:richards}), the HCF should also be taken into account. Ref. \cite{seki2023} shows that some soils, for example, clay soil like UNSODA 2363 (Fig. \ref{fig:unsoda2363}), VG - Mualem model can fit the WRF and the HCF in a wide pressure head, but many soils require a bimodal-CH model with the general HCF. For example, both Gilat loam (Fig. \ref{fig:gilat}) and sand of UNSODA 4661 (Fig. \ref{fig:unsoda4661}) require a bimodal-CH model and the general HCF fitting of the WRF and the HCF. However, as ($h$, $K$) data is not usually available, we may only look at the WRF fitting in these figures. Then Gilat loam requires a DF4 model but UNSODA 4661 only requires a DF3 model. If you are not sure that a DF3 model is sufficient, it would be safe to use a DF4 model for the water flow simulation to give enough flexibility for both water retention and hydraulic conductivity curves.

\section{SWRC Fit}

\subsection{Development of SWRC Fit and unsatfit}

The SWRC Fit is a program for determining the WRF parameters from the measured ($h$, $\theta$) data. It was originally developed with the programming language GNU Octave by using a Levenberg-Marquardt nonlinear regression library (Refs. \cite{seki2007j, seki2007}). Although the Levenberg-Marquardt method requires an initial estimate of the parameter set to be given, which requires some technique to do manually, the SWRC Fit automatically determines a proper set of the initial estimate of the parameters from ($h$, $\theta$) data by a special algorithm explained in Refs. \cite{seki2007, seki2019, seki2023b}. Moreover, a web interface was developed for a simple calculation (Refs. \cite{seki2007, swrcfit}), and you can just give water retention data in a form of the website and the WRF parameters can be obtained with a figure of fitting curves with just one click. Because of its simplicity, the SWRC Fit has been used in many studies; Google Scholar reports that Ref. \cite{seki2007} was cited more than 300 times until 2022. It shows that the algorithm was verified with many types of porous media.

For optimizing the HCF parameters, I started to rewrite code with Python and made a Python library unsatfit. The algorithm for optimizing the WRF parameters was transported from the Octave version of SWRC Fit. It uses a Python library in the SciPy project for nonlinear least-squares optimization with bounds of parameters. It was designed to work with various models (a combination of the WRF and the HCF) as presented in Ref. \cite{seki2022}, with flexible settings of the selecting constant parameters and the upper or the lower bounds of parameters. In addition to determining the WRF and the HCF parameters, the unsatfit can draw a figure like Fig. \ref{fig:unsoda2363}. The website of SWRC Fit was completely rewritten by using the unsatfit library, and the name SWRC Fit only refers to the web interface for determining the WRF parameters now. The source code of the unsatfit and the new SWRC Fit is available at the GitHub repository with an MIT license (Ref. \cite{unsatfitgithub}).

\subsection{How to use SWRC Fit}

\begin{figure}
	\includegraphics[scale=0.65]{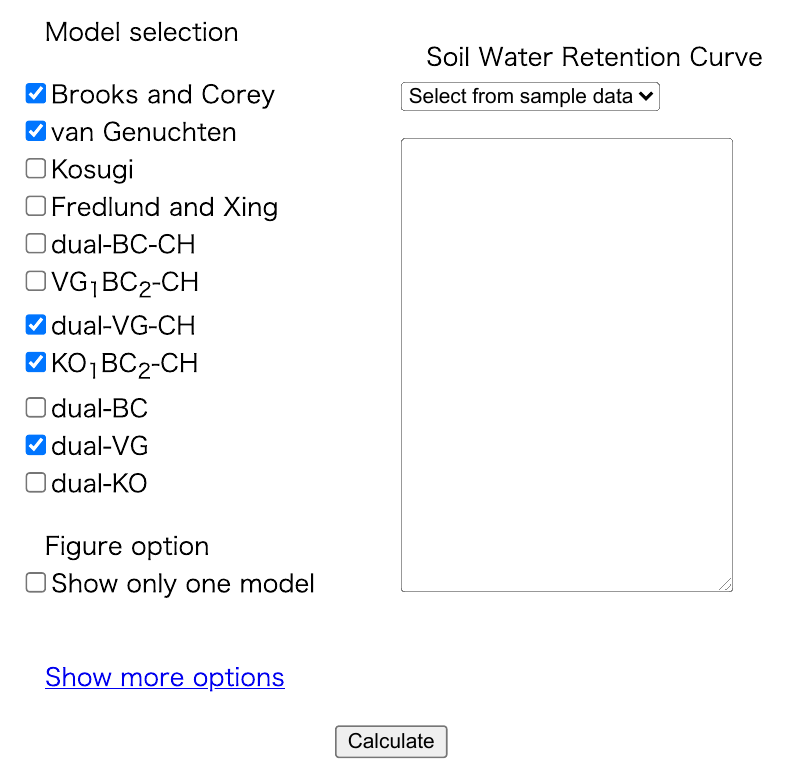}
	\caption{\label{fig:swrcfit}Website of SWRC Fit (Ref. \cite{swrcfit})}
\end{figure}

At the website of the SWRC Fit (Ref. \cite{swrcfit}), the calculation can be conducted as follows. First, copy the soil water retention data ($h$, $\theta$) from a spreadsheet program such as Microsoft Excel, and paste it into the textbox (Fig. \ref{fig:swrcfit}). Sample data can be selected from a pulldown menu. Select models from the "Model selection" menu. Then press the "Calculate" button. Then the name of the model, the equation, the optimized parameters, the coefficient of determination ($R^2$), and the AIC (Akaike information criterion) are listed in a table. The model with the lowest AIC is shown in red color. The measured data and the fitted curves of all the models are shown.

Before calculation, you can check "Show only one model" and then only a fitted curve of a model with the lowest AIC is shown. By clicking "Show more options", setting for the parameter setting of $\theta_s$ and $\theta_r$ can be selected. When the button for calculation is pressed, your selection of models and settings are stored in the web storage (local storage) of your web browser, and it will be effective when you open the page next time.

\section{\label{sec:unsatfit}Unsatfit}

\subsection{Fitting water retention curve}

When only the WRF parameters are needed, usually the SWRC Fit is sufficient. The unsatfit is useful when you want to process many data, or when you want to develop your program by using the function of the parameter determination.

\subsubsection{\label{sec:optimization}Parameter optimization}

The unsatfit package is a library for Python 3 and is distributed with PyPI (Python Package Index). Here follows instruction for an example of parameter optimization of fitting water retention data to the VG equation with the unsatfit.

Install Python 3, and then install the unsatfit and the pandas packages with PyPI by
\begin{lstlisting}
python -m pip install unsatfit pandas
\end{lstlisting}
Prepare a data file in CSV (comma-separated values) format for ($h$, $\theta$) data in the filename “swrc.csv” with a header of “h, theta”. Sample data of UNSODA 2362 (Fig. \ref{fig:unsoda2363}) can be downloaded at the "Sample code" page on the website of the unsatfit (Ref. \cite{unsatfit}).

Now create a text file containing a Python code as follows with a text editor.

\begin{lstlisting}
import numpy as np
import pandas as pd
import unsatfit
ht = pd.read_csv('swrc.csv')
h = np.array(ht['h'])
theta = np.array(ht['theta'])
f = unsatfit.Fit()
f.swrc = (h, theta)
f.set_model('VG', const=['q=1'])
a, m = f.get_init()
f.ini = (max(theta), 0, a, m)
f.optimize()
print(f.message)
\end{lstlisting}

Run this code in the same directory with the file swrc.csv. For running the code on Mac or a Unix-like system, you can add a shebang such as \verb|#!/usr/bin/env python3| at the first line and mark the file executable with the \verb|chmod| command (Ref. \cite{pythonunix}). For running on Windows, please refer to Ref. \cite{pythonwindows}. It then shows optimized parameters as
\begin{quote}
qs = 0.554 qr = 0.000 a = 0.000823 m = 0.101
\end{quote}
\noindent where qs means $\theta_s$, qr means $\theta_r$, and a means $\alpha$ in equation \ref{eq:vg}. Note that the program is unit independent, meaning that the unit of the parameters depends on the unit of the input data. In this case, the unit of the pressure head was cm and therefore the unit of a is cm$^{-1}$, but it is not shown because the program does not know the unit.

Now we look at the sample code step by step.

\begin{lstlisting}
import numpy as np
import pandas as pd
import unsatfit
\end{lstlisting}

These lines load necessary packages; NumPy, pandas, and unsatfit.

\begin{lstlisting}
ht = pd.read_csv('swrc.csv')
h = np.array(ht['h'])
theta = np.array(ht['theta'])
\end{lstlisting}

These lines read the water retention data from swrc.csv using the pandas library. The array of $h$ is stored in the variable h, and the array of $\theta$ is stored in the variable theta, both as NumPy arrays.

\begin{lstlisting}
f = unsatfit.Fit()
\end{lstlisting}

This line defines a variable f as an instance (i.e., object) of the Fit class of the unsatfit library. Let us call the variable f a "fitting object". The unsatfit performs all the calculations by using the methods and the properties defined in the fitting object, which was created from the Fit class. We will see how we can do it hereafter.

\begin{lstlisting}
f.swrc = (h, theta)
\end{lstlisting}

This line sets the property \verb|swrc| of the fitting object as a pair (h, theta). As we have already substituted the water retention data in h and theta, the pair defines the water retention data. Now the fitting object has the water retention data as a property in the name of \verb|swrc|.

\begin{lstlisting}
f.set_model('VG', const=['q=1'])
\end{lstlisting}

This line defines a hydraulic model to be used in the fitting with \verb|set_model()| method defined in the fitting object. The \verb|set_model()| method requires the name of the model as an input parameter. Available models are listed on the "Models" page on the website of the unsatfit (Ref. \cite{unsatfit}). In the van Genuchten (VG) model, "Name" is written as "vg, VG", so we can either use "vg" or "VG" to define the VG model by writing \verb|f.set_model('VG')|. In the list on the website, the WRF parameters are listed as "qs, qr, a, m, q", so we know that $q$ is also a variable. Here we are setting $q=1$ as a constant by \verb|const=['q=1']|. If we also want to make $\theta_r=0$ as a constant, we can write \verb|const=['qr=0', 'q=1']|. If we also want to make $\theta_s=0.554$ as a constant, we can write \verb|const=['qs=0.554', 'qr=0', 'q=1']|. If we want to make $\theta_s$ as the maximum of theta, as calculated by max(theta), then we can write \verb|const=[[1, max(theta)], 'qr=0', 'q=1']|, where \verb|[1, max(theta)]| means that the first parameter in (qs, qr, a, m, q) is equal to max(theta). In this way, we can freely select which WRF parameters are constants and which are free variables to be optimized. As we selected q as a constant parameter, the remaining 4 variables (qs, qr, a, m) are parameters for optimization.

\begin{lstlisting}
a, m = f.get_init()
\end{lstlisting}

This line gets initial estimates of the variables a, m, by using the \verb|get_init()| method of the fitting object. We can see in the VG model at the list on the "Models" page that the \verb|get_init()| method is defined and it returns a, m where q=1. In general, the \verb|get_init()| method returns the WRF parameters without $\theta_s$ and $\theta_r$, where $\theta_s$ is fixed at maximum $\theta$ and $\theta_r$ is fixed at 0. It may not be defined in all models.

Note that as the model was set to VG model by the \verb|set_model()| method, \verb|get_init()| was defined for the VG model, but you can directly use the \verb|get_init_vg()| method to do the same thing without using the \verb|set_model()| method. Actually, the \verb|set_model()| method just defines \verb|get_init()| method as an alias of the pre-defined \verb|get_init_vg()| method.

\begin{lstlisting}
f.ini = (max(theta), 0, a, m)
\end{lstlisting}

This line sets the initial estimate of the parameters to the \verb|ini| property of the fitting object. As the free variables are (qs, qr, a, m), the parameters are set as a vector (tuple or list) in this order, i.e., qs is set to max(theta), qr is set to 0, and a and m are set as the parameters obtained with the \verb|get_init()| method.

\begin{lstlisting}
f.optimize()
\end{lstlisting}

Now optimization with the defined hydraulic model, water retention data, and the initial estimate of parameters is conducted with the \verb|optimize()| method of the fitting object. Parameters are optimized with the nonlinear least-square method, and the result of the optimization is stored in some properties of the fitting object, as shown below.

\begin{lstlisting}
print(f.message)
\end{lstlisting}

It shows a \verb|message| property of the fitting object. It stores the result of the optimization as a text message. When the optimization succeeds, the property \verb|success| is set to True, and the fitted parameters are stored in the \verb|message|. When the optimization fails, \verb|success| is set to False, and the error message is stored in the \verb|message| property. In any case, the result can be shown by running \verb|print(message)|.

Now we finished the explanation of the sample code. In addition, the fitted parameters are stored as a \verb|fitted| property of the fitting object as a list. Therefore, we can obtain the fitted parameters as

\begin{lstlisting}
qs, qr, a, m = f.fitted
\end{lstlisting}

If you want to have $n$, just calculate it by

\begin{lstlisting}
n = 1/(1-m)
\end{lstlisting}

The mean squared error, the standard error, the coefficient of determination, and the AIC are set as properties of the fitting object, as shown in the "Reference" page of the unsatfit website (Ref. \cite{unsatfit}). For example, $R^2$ can be shown by

\begin{lstlisting}
print(f.r2_ht)
\end{lstlisting}
	
\subsubsection{Method of least-square optimization}

The unsatfit uses the least\_squares function for solving a nonlinear least-squares problem in the scipy.optimize module of the SciPy Python package (\verb|scipy.optimize.least_squares|). It uses a trust region reflective algorithm (Ref. \cite{branch1999}) and allows bounds on the variables.

There are several properties of the fitting object which send parameters to the least\_squares function, as shown in the "Properties for settings" section of the "Reference" page of the unsatfit website (Ref. \cite{unsatfit}). Among those properties, \verb|lsq_ftol| is explained here. It is a \verb|ftol| parameter of the least\_squares function, which is a tolerance for the termination concerning $dF/F$ where $F$ is the cost function. The default value of \verb|lsq_ftol| is defined as a list \verb|[0.1, 0.01, 1e-3, 1e-4, 1e-6, 1e-8]|, meaning that \verb|ftol| changes in this order. In this way, if the optimization may fail at some value of \verb|ftol|, the result with the previous \verb|ftol| value can be used.

\subsubsection{Bound of parameters}

The bound of the parameter is listed on the "Models" page of the unsatfit website (Ref. \cite{unsatfit}). Note that the bounds for the HCF parameters are included in this list. For example, the bound of parameter $\alpha$ in the VG model is defined as a property \verb|b_a| as the pair of (minimum, maximum). Most of the default value is set as (0, np.inf), where np.inf means infinity $\infty$. As an exception, b\_m for $m$ in the VG model and b\_w1 for the weighting factor of the bimodal model is set as (0, 1) from the definition.

\subsubsection{Draw fitting curves}

After optimization with the \verb|optimize()| method, a figure for measured data and a fitting curve can be drawn. First, we specify how we present the figure. The \verb|show_fig| property is whether the figure is shown on the screen, and the \verb|save_fig| property is whether we save a figure to a file. Both \verb|show_fig| and \verb|save_fig| are set as False by default. For example, if we want to save a figure file in the png format as 'swrc.png', we set
\begin{lstlisting}
f.save_fig = True
f.filename = 'swrc.png'
\end{lstlisting}
where the \verb|filename| property is for the name of the image file to save. The output format is deduced from the extension of the filename. For example, for obtaining a pdf format file, we can name it 'swrc.pdf'. There are many other properties for drawing figures, such as the size of the figure, the setting for the axis, the plot, the curves, the label, and the legend, as available in the "Properties for settings" section of the "Reference" page of the unsatfit website (Ref. \cite{unsatfit}).

After the setting of the figure is provided, the figure can be drawn with the \verb|plot()| method as
\begin{lstlisting}
f.plot()
\end{lstlisting}

For drawing multiple curves, we can repeat the \verb|add_curve()| method instead of the \verb|plot()| method and use the \verb|plot()| method for the last curve. For clearing the curves, the \verb|clear_curves()| method can be used.

\subsection{Fitting hydraulic conductivity curve}

In the "Sample code" page of the unsatfit website (Ref. \cite{unsatfit}), the codes for optimizing the WRF and the HCF parameters with the same method as Ref. \cite{seki2023} where the optimizing algorithm was verified with various soil samples are provided. Sample data should be prepared as described in the previous section. In addition to the ($h$, $\theta$) data, ($h$, $K$) data should be prepared in the filename “hcc.csv” with a header of “h, K”, where the sample data is available on the page of the sample code.

From here, the sample code for the KBC (KO$_1$BC$_2$-CH) model with $\theta_r=0$ and $r=1$ is described step by step.

\begin{lstlisting}
#!/usr/bin/env python3
import numpy as np
import pandas as pd
import unsatfit
\end{lstlisting}

This is the shebang and import of the packages as described previously.

\begin{lstlisting}
MODEL = 'KBC'
HB = 2
\end{lstlisting}

The parameter MODEL expresses the name of the model, and the parameter HB expresses the $h_b$ value for the modified model.

\begin{lstlisting}
ht = pd.read_csv('swrc.csv')
h_t = np.array(ht['h'])
theta = np.array(ht['theta'])
ht = pd.read_csv('hcc.csv')
h_k = np.array(ht['h'])
k = np.array(ht['K'])
\end{lstlisting}

The WRF and the HCF parameters are read.

\begin{lstlisting}
f = unsatfit.Fit()
\end{lstlisting}

A fitting object is set as a variable f, as explained previously.

\begin{lstlisting}
f.swrc = (h_t, theta)
f.unsat = (h_k, k)
\end{lstlisting}

Now the ($h$, $\theta$) data is the \verb|swrc| property of the fitting object, and the ($h$, $K$) data is stored in the \verb|unsat| property. Note that when \verb|unsat| has data, the optimization will be for both WRF and HCF parameters, while when \verb|unsat| is an empty list (default value), the optimization is only for the WRF parameters.

\begin{lstlisting}[basicstyle=\footnotesize]
qs,qr,w1,hm,s1,l2 = wrf = f.get_wrf_kobcch()
\end{lstlisting}

The \verb|get_wrf_kobcch()| method is a \verb|get_wrf()| for the KBC model, as described on the "Models" page, and it gets all the WRF parameters as a vector. The vector is stored in the parameter wrf for the use of setting a constant parameter set for optimizing the HCF parameters. The respective values of the parameters are stored in the parameters qs, qr, w1, hm, s1, l2 successively, where s1 for $\sigma_1$ will be used later.

\begin{lstlisting}
model = MODEL
\end{lstlisting}

The parameter model is again defined for the name of the model, as the model name may change to "MKBC" for the modified KBC model.

\begin{lstlisting}
f.set_model(model, const=[wrf, 'r=1'])
\end{lstlisting}

As the parameter model is 'KBC', the KBC model is defined with a constant parameter set of \verb|[wrf, 'r=1']|. As the fitting object has the HCF parameters in \verb|unsat| property and it is now working as an HCF optimization mode, the parameters in the model also include the HCF parameters. In the "Models" page, "parameters which only appear in HCF" is listed as Ks, p, q, r. For most of the models, The HCF parameters are appended at the end of the WRF parameters, and therefore all the parameters are listed as qs, qr, w1, hm, sigma1, l2, Ks, p, q, r. An exception is for the models with the VG function or subfunction, where the WRF parameter q comes between p and r. When a vector longer than 2 elements is provided, it is regarded as a list of the WRF parameter set. In this case, as wrf was given, all the WRF parameters are set as constant parameters. In addition, 'r=1' is given, and therefore r is set as a constant parameter. The remaining parameters, Ks, p, and q are the parameters for optimization.

\begin{lstlisting}
if s1 > 2:
    model = 'M' + MODEL
    f.modified_model(HB)
\end{lstlisting}

The \verb|modified_model()| method changes the defined model to the modified model with equation \ref{eq:modified}, where the $h_b$ value should be given as an input variable. When $\sigma_1 > 2$, a modified model is used. The name of the model is changed to "MKBC", and the modified model with $h_b$ value as the HB parameter initially defined is set. It is noted that for producing a modified model, the WRF parameters should start with (qs, qr) and the HCF parameter should start with Ks, as the method uses the information of (qs, qr, Ks) from the location of the parameters.

\begin{lstlisting}
print(f.model_description)
\end{lstlisting}

The \verb|model_description| stores the information on the name of the model and the values of constant parameters. For example, when you run this program with UNSODA 2363 sample data, this line prints as follows.
\begin{quote}
Modified KBC model (hs = 2) with qs = 0.554 qr = 0.000 w1 = 0.397 hm = 7999.99 sigma1 = 2.005 l2 = 0.0148 r = 1.000
\end{quote}
where hs is the $h_b$ value.

\begin{lstlisting}
p = (1, 2, 4, 6)
q = (0.5, 1, 2)
f.ini = ((max(k),), p, q)
\end{lstlisting}

The initial parameter set is defined here. The initial value of $K_s$ is defined as the maximum value of the measured $K$ value. The initial parameters for $p$ and $q$ are given as vectors. When the initial parameter is given as a vector, all the parameters in the vector are set as the initial parameter. This is the following algorithm as described in Ref. \cite{seki2023}.
\begin{quote}
The HCF parameters were optimized in two steps. At first, multiple initial conditions, notably (1, 2, 4, 6) for $p$ and (0.5, 1, 2) for $q$ or $r$ (12 combinations in total) were used and the parameters optimized with a relatively loose convergence criterion to facilitate rapid calculations. The fitted parameter set with the least MSE was used next as initial condition for a second optimization step using much stricter convergence criteria to obtain more accurate parameter values.
\end{quote}
In this citation, "a relatively loose convergence criterion" refers to ftol value as described before. It is stored in \verb|lsq_ftol_global| property of the fitting object as [1, 0.1]. Therefore, all the 12 combinations of the initial parameters are optimized to ftol=0.1 first and then the parameter set with the least MSE is used for the initial condition for the second optimization to ftol=$10^{-8}$ as normal. This is a key technical tip that led to success in the optimization of various soils in Ref. \cite{seki2023}.

\begin{lstlisting}
f.b_p = (0.3, 10)
f.b_q = (0.1, 2.5)
\end{lstlisting}

These lines define the bounds of $p$ and $q$. They could be adjusted if desired.

\begin{lstlisting}
max_k = max(k) * 2
if min(h_k) > 1:
    max_k = max_k * (min(h_k)) ** 2
f.b_ks = (max(k) * 0.95, max_k)
\end{lstlisting}

These lines define the bounds of $K_s$. When the minimum value of $h$ for the ($h$, $K$) dataset is large, uncertainty is large and the upper limit is set higher. This is unit dependent on $h$, and therefore it could be adjusted.

\begin{lstlisting}
f.optimize()
\end{lstlisting}

Now everything is ready and we optimize the parameters.

\begin{lstlisting}
if not f.success:
    print(f.message)
    exit(1)
\end{lstlisting}

If not successful, show the error message and exit the program.

\begin{lstlisting}
ks, p, q = f.fitted
\end{lstlisting}

It gets the fitted parameters. The values of $p$ and $q$ will be used as a legend in the figure.

\begin{lstlisting}
print('HCF parameters and R2')
print(f.message)
\end{lstlisting}

Now show the fitted parameters as a result.

We omit here several lines which set figure properties. Please check the source of the sample code and "Reference" page of the unsatfit website.

\begin{lstlisting}
f.save_fig = True
f.filename = MODEL + '.png'
f.plot()
\end{lstlisting}

The figure is stored in the filename of "KBC.png".

\section{Other use cases}

\subsection{Direct calculation}

After \verb|set_model()|, the \verb|f_ht(p, x)| function is defined as $\theta(h)$ with the free parameters p (vector) and the pressure heads x. For example, if you want to have $\theta(0.1)$ with the optimized WRF, you can write the code as

\begin{lstlisting}
print(f.f_ht(f.fitted, 0.1))
\end{lstlisting}

For calculating the multiple pressure heads, x can be a NumPy array. In the same way, the \verb|f_hk(p, x)| function is defined as $K(h)$ with the free parameters p (vector) and the pressure heads x.

\subsection{\label{sec:countour}Contour plot}

\begin{figure}
	\includegraphics[scale=0.8]{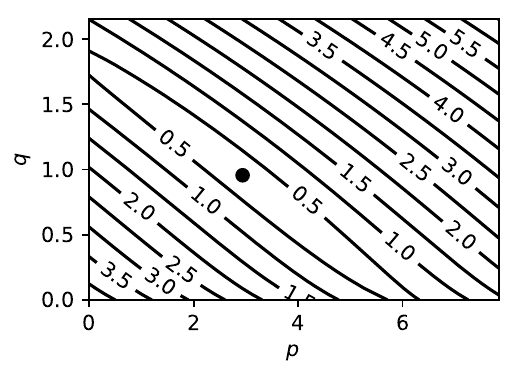}
	\caption{\label{fig:contour}Contour plot of RMSE of log$_{10}(K)$ using various $p$ and $q$ values, and the optimized parameter (closed circle) obtained with the KO$_1$BC$_2$-CH model ($\theta_r=0$) and general HCF ($r=1$) for the sand of UNSODA 4661.}
\end{figure}

A sample code for drawing a contour plot is available on the "Sample code" page of the website of the unsatfit ({Ref. \cite{unsatfit}}). It has some options to set the range of the plot, and for example the contour plot of the parameters $p$ and $q$ can be saved to contour.png as Fig. \ref{fig:contour}. Fig. \ref{fig:contour} is example output for UNSODA 4661 of Fig. \ref{fig:unsoda4661}.

\begin{lstlisting}
f.save_fig = True
f.filename = 'contour.png'
f.contour('p', 'q')
\end{lstlisting}

\bibliography{ref}
\bibliographystyle{apalike}

\end{document}